\documentclass[12pt]{iopart}

\usepackage{graphicx}
\usepackage{citesort}
\usepackage{gensymb}
\usepackage{wasysym}

\begin{document}

\title[PIC Simulations of the Weibel Instability]
{PIC simulations of the Thermal Anisotropy-Driven Weibel Instability: 
Field growth and phase space evolution upon saturation}
\author{A Stockem, M E Dieckmann and R Schlickeiser}
\address{Institute of Theoretical Physics IV, Faculty of Physics and 
Astronomy, Ruhr-University Bochum, D-44780 Bochum, Germany}
\ead{anne@tp4.rub.de}

\begin{abstract}
The Weibel instability is investigated with PIC simulations of an
initially unmagnetized and spatially uniform electron plasma. This
instability, which is driven by the thermally anisotropic electron
distribution, generates electromagnetic waves with wave vectors
perpendicular to the direction of the higher temperature. Two
simulations are performed: A 2D simulation, with a simulation plane
that includes the direction of higher temperature, demonstrates that
the wave spectrum is initially confined to one dimension. The electric
field components in the simulation plane generated by the instability
equalize at the end of the simulation through a secondary
instability. A 1D PIC simulation with a high resolution, where the
simulation box is aligned with the wave vectors of the growing waves,
reveals details of the electron phase space distribution and permits a
comparison of the magnetic and electric fields when the instability
saturates. It is shown that the electrostatic field is driven by the
magnetic pressure gradient and that it and the magnetic field
redistribute the electrons in space.
\end{abstract}


\maketitle

\section{Introduction}\label{intro}
The filamentation instability or beam-Weibel instability has been
proposed as a mechanism to magnetize the early universe
\cite{gru,oka,sch1,sch2,fuj,med1}, to provide the strong magnetic
fields for the afterglow emission of gamma-ray bursts
\cite{ach1,med2,wax,spit} and supernova explosions \cite{med3} and to
explain the heating processes in the pulsar winds \cite{yan,aro1}. 
It is also important for the inertial confinement fusion in 
laser-plasmas \cite{tab}. The filamentation instability has been 
investigated analytically \cite{yoo,ach1,pet1,tau1}, numerically 
\cite{sto3,sil} and also in a laboratory experiment \cite{med3}, 
where it has been discussed as the precursor to the formation of 
astrophysical shocks of gamma-ray bursts and supernovae.

However, the filamentation instability is not the only instability
that can generate magnetic fields. Here, we focus on a second one, the
thermal anisotropy-driven Weibel instability (TAWI), which we study
with particle-in-cell simulations. There are several analytical and
numerical works of the TAWI, e.\,g.\ \cite{aro1,sr1,sch3,mor,rom,puk}.
We illustrate the physical mechanism behind the TAWI for a plasma, in
which the temperature along one direction exceeds significantly those
along the other two directions. The typical microcurrents of the
electrons are highest along the direction with the high
temperature. The magnetic interaction force that causes the deflection
of electrons is thus strong orthogonal to this direction. At the same
time, the typical particle's thermal energy orthogonal to the
direction with the high temperature is relatively weak. The particle's
thermal motion can, in this case, not oppose the structure
formation. Current channels form with an axis, which is aligned with
the direction of the high plasma temperature.

The original work of Weibel \cite{wei} considers a bi-Maxwellian
distribution with one component of lower temperature yielding a one
dimensional wave growth with wave vectors \(\mathbf k\) in the
direction of lower temperature. In contrast, we choose only one hot
direction. The wave growth will then occur in the two dimensional 
plane perpendicular to that direction and give rise to the generation
of a magnetic field in this plane.

The temperature anisotropy we consider here can be the result of a
Buneman instability \cite{ish}, which emerges for example at the
terrestrial foreshock \cite{eas1}. The latter is a result of the
backstreaming of solar wind particles, which do not pass the
terrestrial bow shock and move back into the solar wind domain. The
backstreaming particles form a foreshock upstream of the shock. A
Buneman instability develops between the shock-reflected ions and the
electrons of the solar wind, if we assume for simplicity that no
magnetic field is present. The phase velocity of the waves is
comparable to the ion beam velocity \cite{bun}, which we denote here
as \(v_b\) in the electron rest frame. The electric field amplitude
grows and the electrons are trapped in phase space holes, when the
kinetic energy of the electrons in the wave frame is not sufficient to
escape the wave potential \cite{ber,die5}. They are accelerated along
the beam velocity vector to a few times the beam speed. Since the
spectrum of the waves driven by the Buneman instability is not
strictly mono-directional, the electrons also experience a, usually
weaker, acceleration away from the beam velocity vector. After the
heating of the electrons, the Buneman instability is quenched. The
electrons are heated more in the ion beam direction than in the plane
orthogonal to it and a temperature anisotropy has developed.

The proximity of the terrestrial foreshock permits a direct access by
in-situ observations. Therefore the particle configurations are
well-known and the ion beam velocity could be determined to some
hundred km\(/\)s \cite{eas1}, which is comparable to the speed of the
Solar wind in the reference frame of the bow shock.  We assume that
this physical scenario can be applied to the foreshocks of supernova
remnants (SNR), where the bulk flow speed of the supernova blast shell
and thus of the SNR shock has been determined to be less or equal to
\(0.2 c\) \cite{kul}. The ion beam speed should be similar to the
velocity of the shock in the upstream reference frame.

Realistic values for the interstellar medium, into which a SNR shock
is expanding, would be \(B_{ISM} = 1 \) nT and \(n_{ISM} \approx 1
\textnormal{ cm}^{-3}\) \cite{ellison}, which would give a ratio of
the plasma frequency to the electron cyclotron frequency of \(\approx
300\). Thus, we can neglect the magnetic fields. Experimental evidence
indicates that the magnetic field close to SNR shocks is amplified to
values \(\approx 100\) nT, presumably by the shock-accelerated cosmic
ray particles \cite{volk,bell}. Recent PIC simulations \cite{niemiec}
seem to suggest though, that the magnetic field cannot be amplified by
the cosmic rays to values much higher than the amplitude of the
background magnetic field.

The TAWI we consider here is driven by the electron heating by the
shock-reflected ion beam, which we find even at unmagnetized shocks
\cite{forslund}.  It can thus amplify a magnetic field from
fluctuation levels and constitute the first stage of a chain of
magnetic field amplification mechanisms upstream of a SNR shock. Our
PIC simulations evidence that the peak magnetic field, which results
from the TAWI, would be 8 nT for a plasma density of \(1 \textnormal{
  cm}^{-3}\), well in excess of the ISM magnetic field. These 8 nT
constitute a minimum estimate, because the electron density, which
scales this peak field, is higher in the foreshock than in the ISM.

We perform two simulations and compare them to \cite{mor}: First we
choose a 2D simulation plane that includes the direction of higher
temperature, which we refer to as the parallel direction, and one
perpendicular direction with a lower temperature. We show that the
wave spectrum is one-dimensional during the early stage of the
instability. In reality, this wave spectrum would be spread out over
the plane that is orthogonal to the parallel direction. The
dimensional reduction is, however, beneficial for the study of the
interplay of the electromagnetic fields and the key plasma
structures. We discuss the temporal behaviour of the electric and
magnetic energy densities and analyse the field data, demonstrating
that only that component of the magnetic field is amplified, which
points out of the simulation plane. The magnetic pressure gradient
generates initially an electric field in the perpendicular direction.
A secondary instability transfers electric field energy from the
perpendicular into the parallel direction.

Second we perform a 1D simulation along the perpendicular
direction. We demonstrate that a weak electric field in the parallel
direction is generated by Ampere's law. The connection between the
much stronger electric field in the perpendicular direction and the
magnetic pressure gradient is demonstrated also in this setting. The
growing magnetic field pushes the electrons together, yielding an
increasingly complex layered structure in the velocity distribution.

The paper is structured as follows. The initial conditions and the 
simulation setup are given in the section \ref{incon}. The results of 
the 2D and 1D simulations are presented in the sections \ref{zweiD} and 
\ref{einD}, respectively. Section \ref{discussion} discusses the results 
and the future work.

\section{The instability, the initial conditions and the simulation 
method}\label{incon}

\subsection{The linear instability}

We discuss the TAWI for electrons in a homogeneous, collisionless
plasma and leave the ions as neutralising immobile background, which
is equivalent to an infinitely large ion mass (\(m_i \rightarrow
\infty\)). The plasma is initially unmagnetized (\(B_0=0\)). A
temperature anisotropy \(A=(v_{th\parallel} / v_{th\perp})^2-1 >0\) in
the electron particle distribution
\begin{equation}
	f_0(v_\perp, v_\parallel) = \frac{1}{\pi^{3/2}c^3
	v_{th\perp}^2 v_{th\parallel}} \exp
	\left[-\left(\frac{v_\perp^2}{v_{th\perp}^2}+\frac{v_\parallel^2}{v_{th\parallel}^2}\right)\right]
\end{equation}
leads to growing electromagnetic oscillations with a wave vector \(\mathbf k\) perpendicular to the direction of higher temperature. The velocities are normalized to the speed of light \(c\).

We denote the direction of high temperature as the \(x\)-direction
(\(v_{th\parallel} = v_{thx}\)). Hence, the wave growth of the TAWI
mode can occur in the \(y\)-\(z\) plane. We can limit our analysis to the 
\(y\)-direction due to the isotropy in this plane 
(\(v_{th\perp}=v_{thy}\)), accordingly
\(v_{th\perp}<v_{th\parallel}\).

The chosen thermal velocities are \(v_{th\perp}=8.83 \cdot 10^{-3}\)
and \(v_{th\parallel}=8.83 \cdot 10^{-2}\), yielding \(A=99\).
The parallel thermal velocity is a few percent of the speed of light,
which matches with the ion beam speed upstream of the SNR shocks. The
perpendicular electron thermal velocity corresponds to a thermal
energy of 40 eV.

The dispersion relation for the linear phase of the instability
\cite{mar1,mor}
\begin{equation}
	k^2 + \sigma^2 = - \left[1+\frac{1}{2} (A+1) Z'\left(\frac{\imath \sigma}{k v_{th\perp}}\right)\right]
\end{equation}
describes the growth of the electromagnetic oscillations with a wave
number \(k\) in units of the inverse electron skin depth \(\omega_{p}
/ c\) and its associated linear growth rate \(\sigma\) in units of
\(\omega_{p}\). The frequency is purely imaginary (\(\omega=\imath
\sigma\)), thus the oscillations do not propagate in space and grow
exponentially in time. The electron plasma frequency for the number
density $n_e$ is given by \(\omega_{p}=(e^2 n_e/ \epsilon_0
m_e)^{1/2}\) and \(Z'(\zeta)=-2[1+\zeta Z(\zeta)]\) is the first
derivative of the plasma dispersion function \(Z(\zeta)= \pi^{-1/2}
\int_{-\infty}^\infty d t \, \exp(-t^2)/(t-\zeta) \).

\begin{figure}
\centering
  \setlength{\unitlength}{0.001\textwidth}
  \begin{picture}(1100,450)(-260,0)
    \includegraphics[width=7cm]{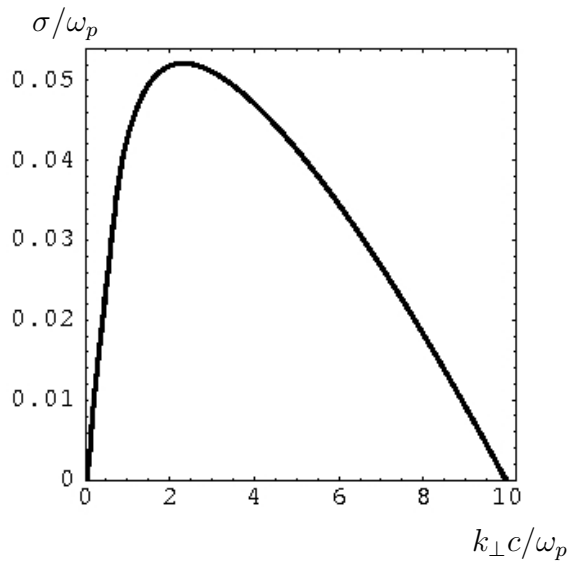}
    \put(-50,-8){\(k_\perp c/\omega_p\)}
    \put(-420,430){\(\sigma/\omega_p\)}
  \end{picture}
\caption{The growth rate \(\sigma(k)\). After a steep rise at low wave
  numbers the maximum is reached. Thermal effects limit the growth
  rate at large \(k\). The wave number cutoff is given by
  \(A^{1/2}\).}\label{Plot1}
\end{figure}

The plasma system is unstable for wave numbers in the range \(0< k <
k_{max} \) as depicted by the Fig.\ \ref{Plot1}, where
\(k_{max}=A^{1/2}\) depends on the temperature ratio \(T_\parallel /
T_\perp >1\). The instability generates a magnetic field in
perpendicular direction \(B=B_\perp = (B_y^2+B_z^2)^{1/2}\) and an
electric field in parallel direction \(E=E_\parallel = E_x\). The
connection between \(E\) and \(B\) is given by the Maxwell relation
\(\mathbf k \times \mathbf E = \omega \, c \, \mathbf B\).

\subsection{The simulation method and parameters}
The PIC simulation is a method to model self-consistently the
interplay of the electric and magnetic fields with the particles of a
collision-less kinetic plasma \cite{daw2}. The plasma is treated as an
incompressible phase space fluid, which is approximated by an ensemble
of volume elements. The Vlasov equation is solved by the method of
characteristics, i.\,e.\ a Lorentz equation of motion is solved for
each volume element. A volume element contains a large number of
physical particles, which are treated as one computational particle
(CP) with the same charge to mass ratio as the physical particles. The
velocity of a CP is \(\mathbf{v}_{cp}\) at the position
\(\mathbf{x}_{cp}\). With the relativistic momentum
\(\mathbf{p}_{cp}=m_{cp}\gamma \mathbf{v}_{cp}\) that is given in
terms of \(m_e c\) and \(m_{cp}\) in terms of the electron mass
\(m_e\), the normalized Maxwell equations for the electric and
magnetic fields \(\mathbf{E}(\mathbf{x},t)\) and
\(\mathbf{B}(\mathbf{x},t)\)
\begin{equation}
	\nabla\times \mathbf E=- \frac{\partial \mathbf B}{\partial t}, \, \,  \nabla \times \mathbf B=  \frac{\partial \mathbf E}{\partial t}+\mathbf J,    \label{thirteen}
\end{equation}
and the normalized Lorentz equation
\begin{equation}
	\frac{\textnormal{d}\mathbf p_{cp}}{\textnormal{d} t}=q_{cp}(\mathbf E [\mathbf{x}_{cp}]+\mathbf v_{cp} \times   \mathbf B[\mathbf{x}_{cp}]) \label{fourteen}
\end{equation}
are solved. The quantities \(\mathbf E\) and \( \mathbf{B} \) can be 
transformed to physical units by a multiplication by \(\omega_{p} m_e /c e \) 
and \(\omega_{p} m_e /e \), respectively. The normalized current density 
\( \mathbf{J} \) is transformed to physical units by a multiplication with 
\( c n_e e\). The \(q_{cp}\) is given in units of the elementary charge \(e\) 
and the simulation time \(t\) and position \(\mathbf{x}\) are given in units 
of \(1/\omega_p\) and \(c/\omega_p\), respectively. The code fulfills 
\(\nabla \cdot \mathbf E=\rho/\epsilon_0\) and \(\nabla \cdot \mathbf B=0\) 
to round-off precision \cite{eas2}.

The current $\mathbf{J}$, the electric and the magnetic fields are
defined on a grid. The electromagnetic fields are interpolated from
the grid to the position of each CP. The momentum of the CP is updated
with these interpolated fields and with the Lorentz equation of
motion. The new particle position is calculated with the old particle
position, \(\mathbf{v}_{cp}\) and the time step \(\Delta_t\). The
microcurrents \( \propto q_{cp} \mathbf{v}_{cp}\) are the
contributions of each discrete particle, i.\,e.\ the CPs in our
simulations or the electrons in a real physical plasma. They are
interpolated from each CP back onto the grid and by integrating over
all CPs the total current \(\mathbf{J}\) is obtained. Then the
electric and magnetic fields are updated with (\ref{thirteen}) and the
individual steps are repeated.

Our PIC simulations represent either one or two spatial dimensions and 
all components of \( \mathbf{p}_{cp}\). The boundary conditions are 
periodic in all directions. The quadratic 2D simulation box employs an 
equal number of cells \(N_x=N_y=1000\) in the \(x\)- and in the 
\(y\)-direction, each with a side length \(\Delta_x \omega_p / c = 8.4
\times 10^{-3}\).  The number of CPs per cell is \(N_2= 150\). For the 
1D simulation we use \(N_y= 1000\) cells in the \(y\)-direction, \(N_x=1\) 
and the same \(\Delta_x \). The number of CPs per cell is in this case 
\(N_1=32768\). The time step is given by \(\Delta_t \omega_p \approx 3.9 
\cdot 10^{-3}\). The equations solved by the PIC code are given in 
normalized units and they can be scaled to any value of the plasma 
frequency. We choose a reference value for the plasma frequency 
(\(\omega_p = 6.3 \cdot 10^5 \textnormal{ s}^{-1}\)) to illustrate the 
field amplitudes in physical units.

\section{Two-dimensional simulation}\label{zweiD}

\subsection{The electric and magnetic energy densities}
The box-averaged energy densities of the electric $E_i (x,y,t)$ and
magnetic $B_i (x,y,t)$ field components are given by \(\epsilon_{Ei}
(t) =(N_x\, N_y)^{-1} \sum \limits_{j,k} \epsilon_0 {\left [
    E_i(j\Delta_x,k\Delta_x,t) \right]}^{2}/2\) and
\(\epsilon_{Bi}(t)= (N_x\, N_y)^{-1} \sum \limits_{j,k} {\left [
    B_i(j\Delta_x,k\Delta_x,t) \right ]}^2 /2 \mu_0\), respectively.
The box-averaged kinetic energy density is \(\epsilon_{K}(t)=(N_x\,
N_y)^{-1} \Delta_x^{-3}\sum \limits_j m_{cp} \, c^2(\gamma_j-1)\),
where the summation is over the \(N_y \,N_1\) or \(N_x N_y  \, N_2\)
particles for the 1D or the 2D simulation, respectively.
The only magnetic field component that grows in the considered
geometry in response to the TAWI is the \(B_z\) component, which is
confirmed by the PIC simulations.

\begin{figure}[!ht]
	\centering \includegraphics[width=7cm]{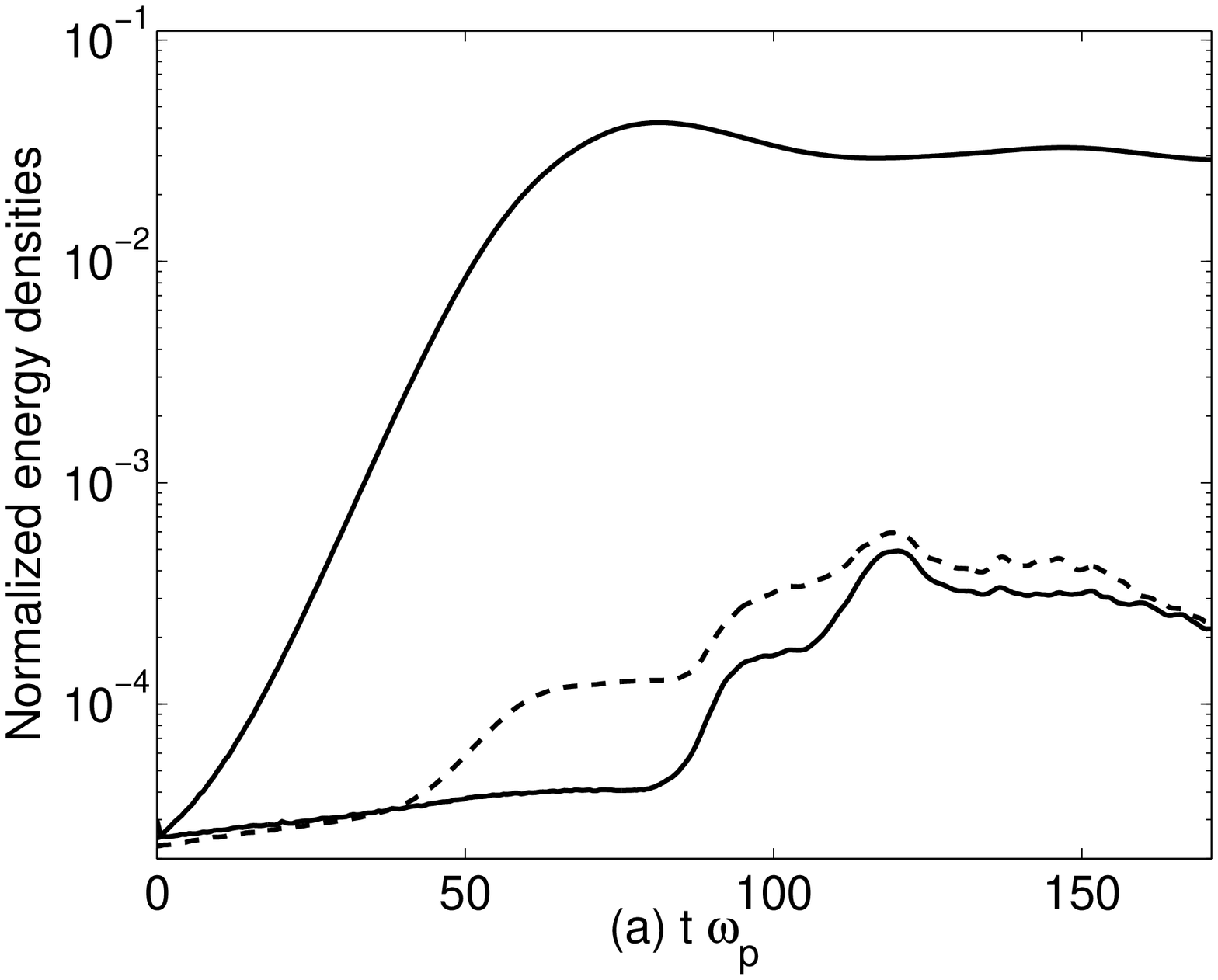}
	\includegraphics[width=7cm]{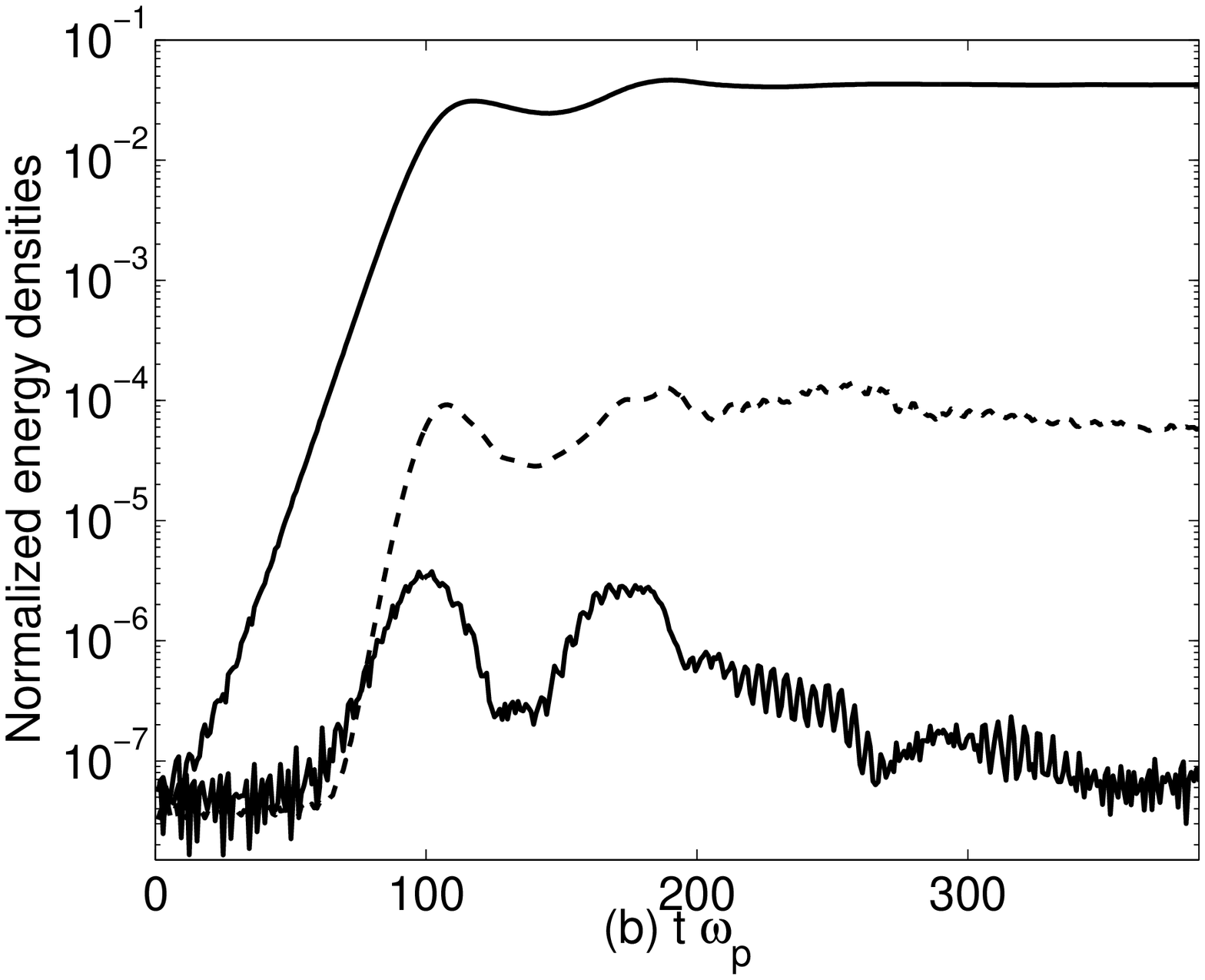}
	\caption{10-logarithmic plot of the ratios
          \(\epsilon_{Bz}/\epsilon_K(t=0)\) (upper solid),
          \(\epsilon_{Ey}/\epsilon_K(t=0)\) (dashed) and
          \(\epsilon_{Ex}/\epsilon_K(t=0)\) (lower solid) for the 2D
          (a) and 1D simulation (b). The TAWI amplifies \(B_z\), which
          induces a component \(E_y\). In 2D \(E_x\) is amplified
          later than \(E_y\) and their energies eventually
          equalize.}\label{fig:S1enden}
\end{figure}

Fig.\ \ref{fig:S1enden} demonstrates that the energy density of
\(B_z\) increases exponentially first and saturates for large
simulation times similar to \cite{mor}. The electric field component
\(E_y\) is also amplified. The energy density of \(E_y\) does not
grow in the 2D box at twice the exponential rate, with which the
energy density of \(B_z\) is growing. It does so only in the 1D
simulation (shown here for comparison, Fig.\ \ref{fig:S1enden}(b)).
The growth of the component \(E_x\) is delayed with respect to
\(E_y\). Eventually, both components \(E_x\) and \(E_y\) reach the
same energy density in Fig. \ref{fig:S1enden}(a). In reference
\cite{mor} the electric energy density is just a horizontal line
because the energy densities have been plotted on a linear scale. The
energy densities of the other field components did not grow during the
simulation time.

\subsection{The field evolution}
\begin{figure}[!ht]
	\centering \includegraphics[width=7cm]{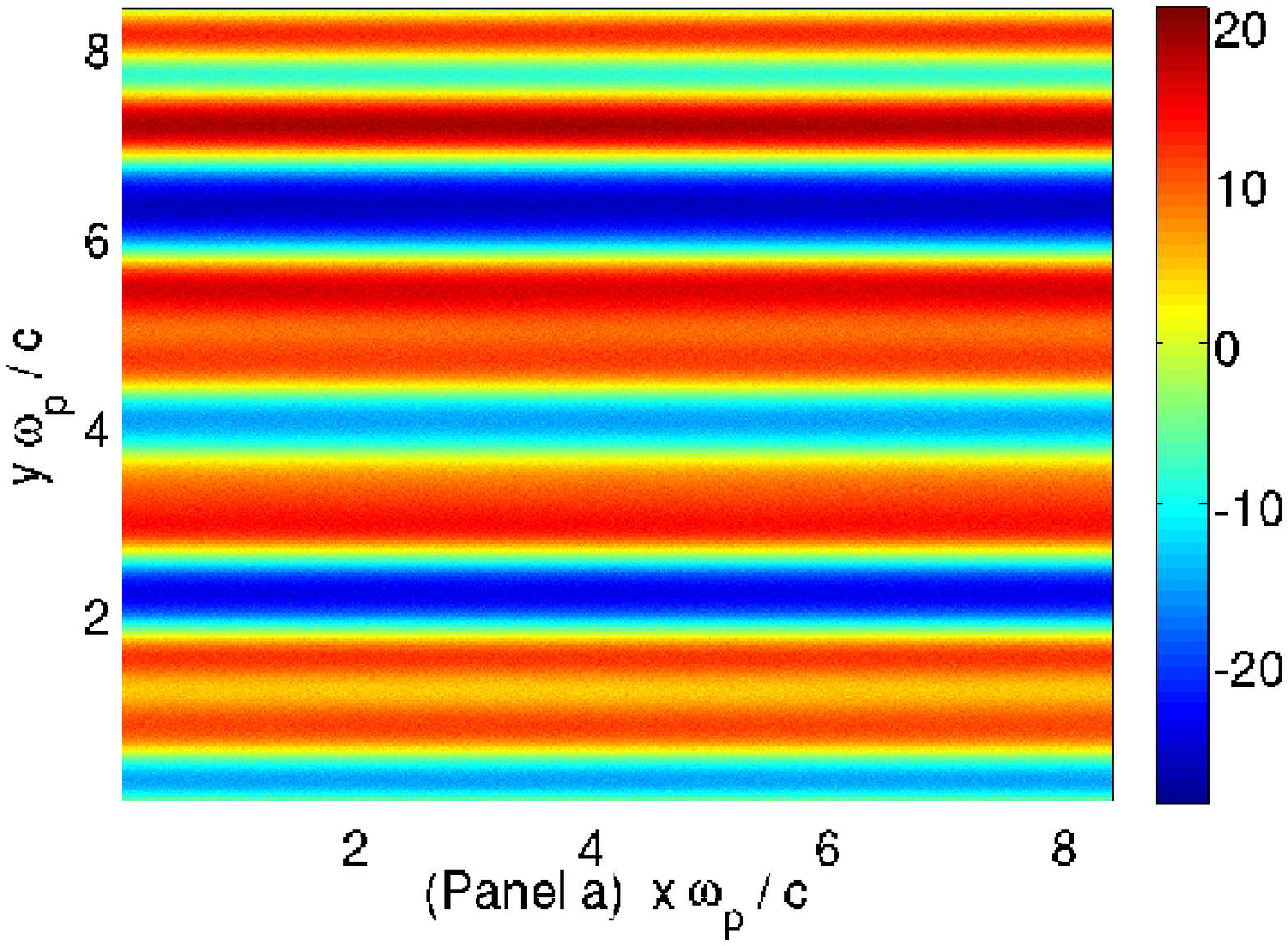}
	\includegraphics[width=7cm]{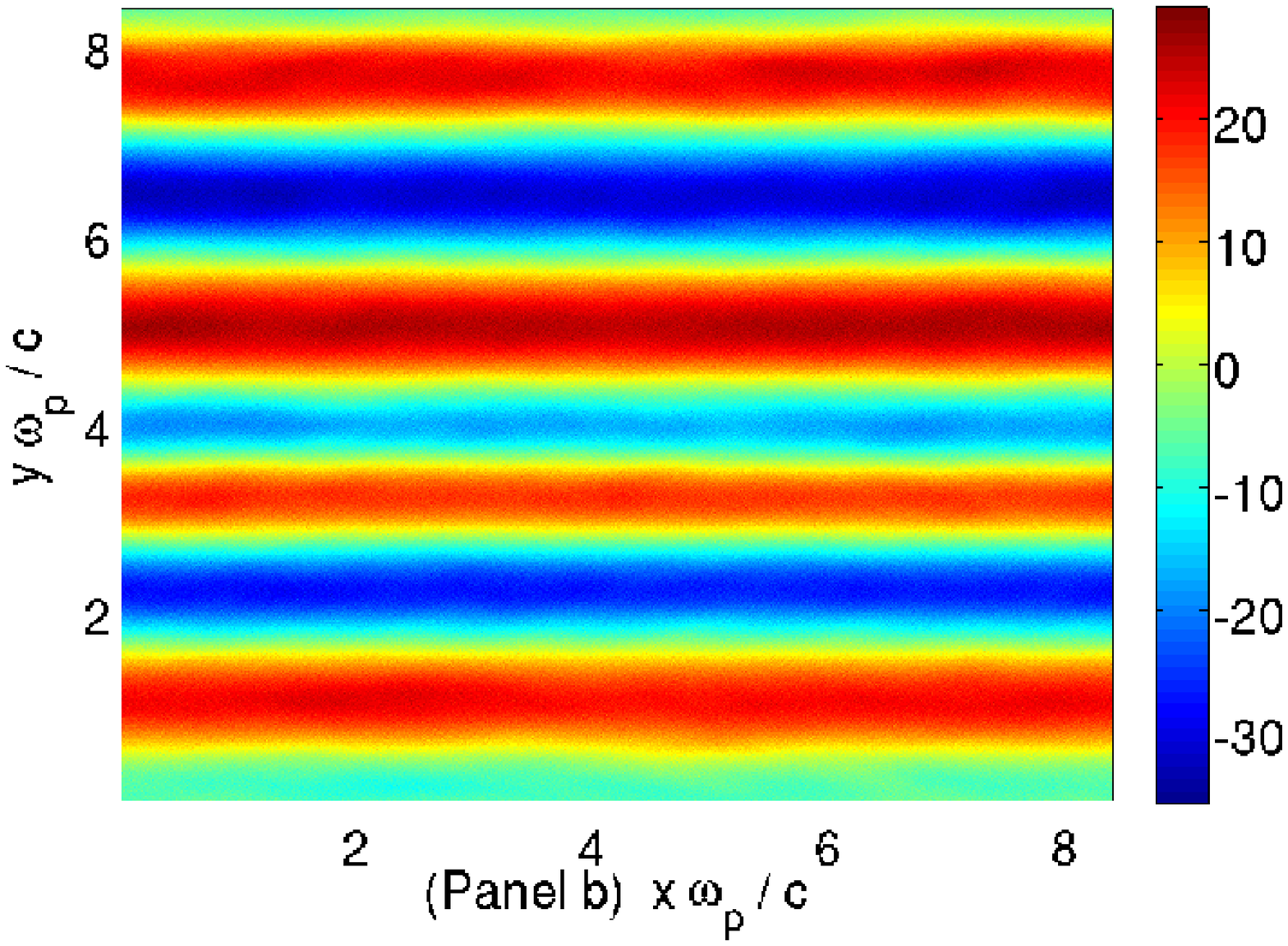} \caption{The magnetic
	field component \(cB_z\) at the simulation times \(t \omega_p
	= 59\) (a) and \(t \omega_p = 164\) (b). The colourbar denotes
	the value in V/m (colours online). After the onset of the
	amplification of \(E_x\) the filaments are not strictly
	one-dimensional anymore.}\label{fig:S1BZ120}
\end{figure}
Fig.\ \ref{fig:S1BZ120} displays the magnetic field component \(cB_z\)
at the simulation times \(t \omega_p =59\) and \(t \omega_p =
164\). The left panel shows the magnetic field component \(cB_z\) at
the time when \(E_y\) saturates first. The component \(E_x\) has not
been amplified yet. Magnetic filaments have developed with a symmetry
axis parallel to the \(x\)-direction. The right panel displays the
magnetic field component when both electric field components have
reached almost the same amplification level. The magnetic field
structures are not strictly parallel to the \(x\)-axis anymore
(e.\,g.\ at \(y \omega_p / c= 6.5\)), but they remain qualitatively
unchanged in the considered reduced geometry. The structures would
merge if the two perpendicular directions would be resolved
\cite{mor}.

\begin{figure}[!ht]
	\centering \includegraphics[width=7cm]{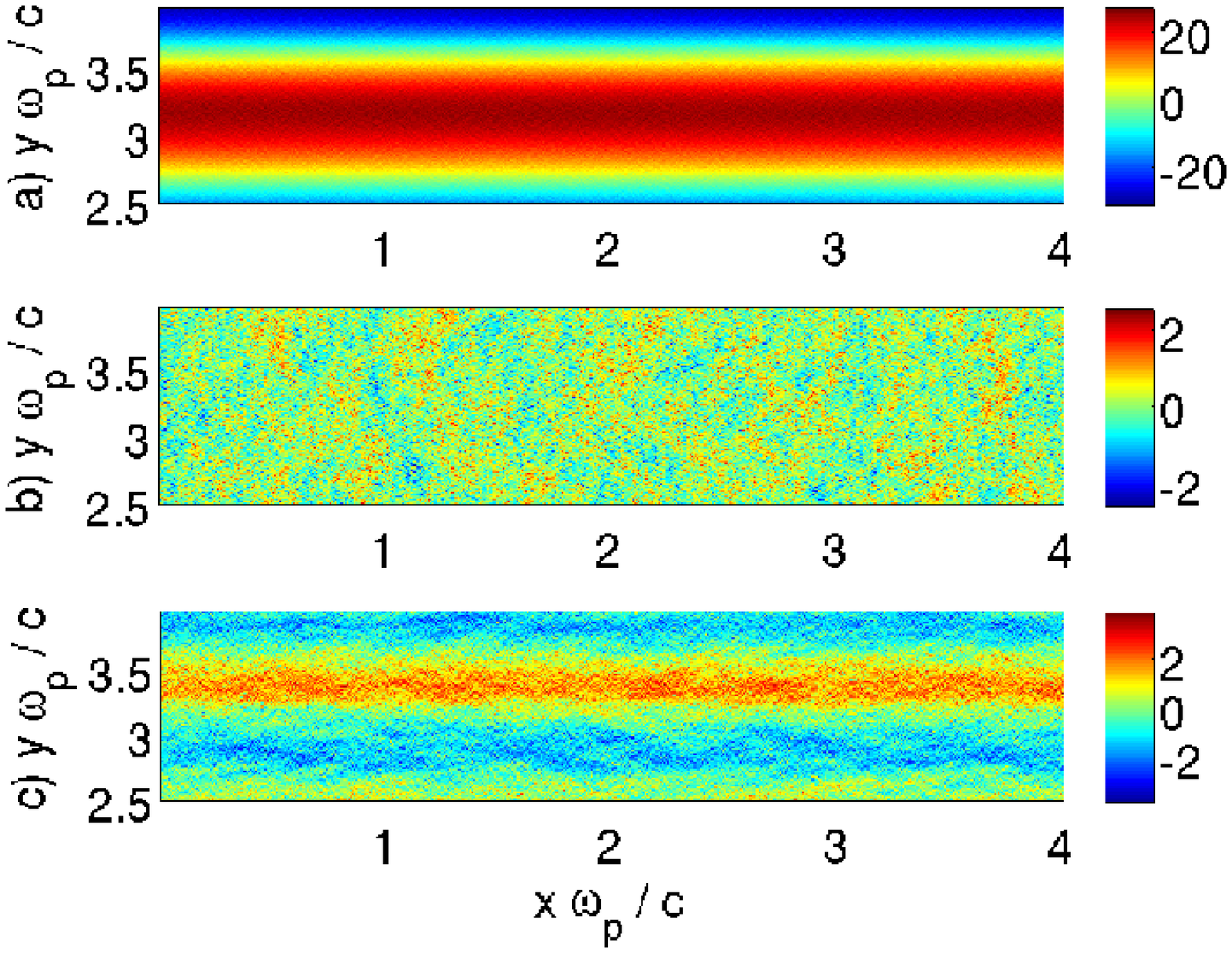}
	\includegraphics[width=7cm]{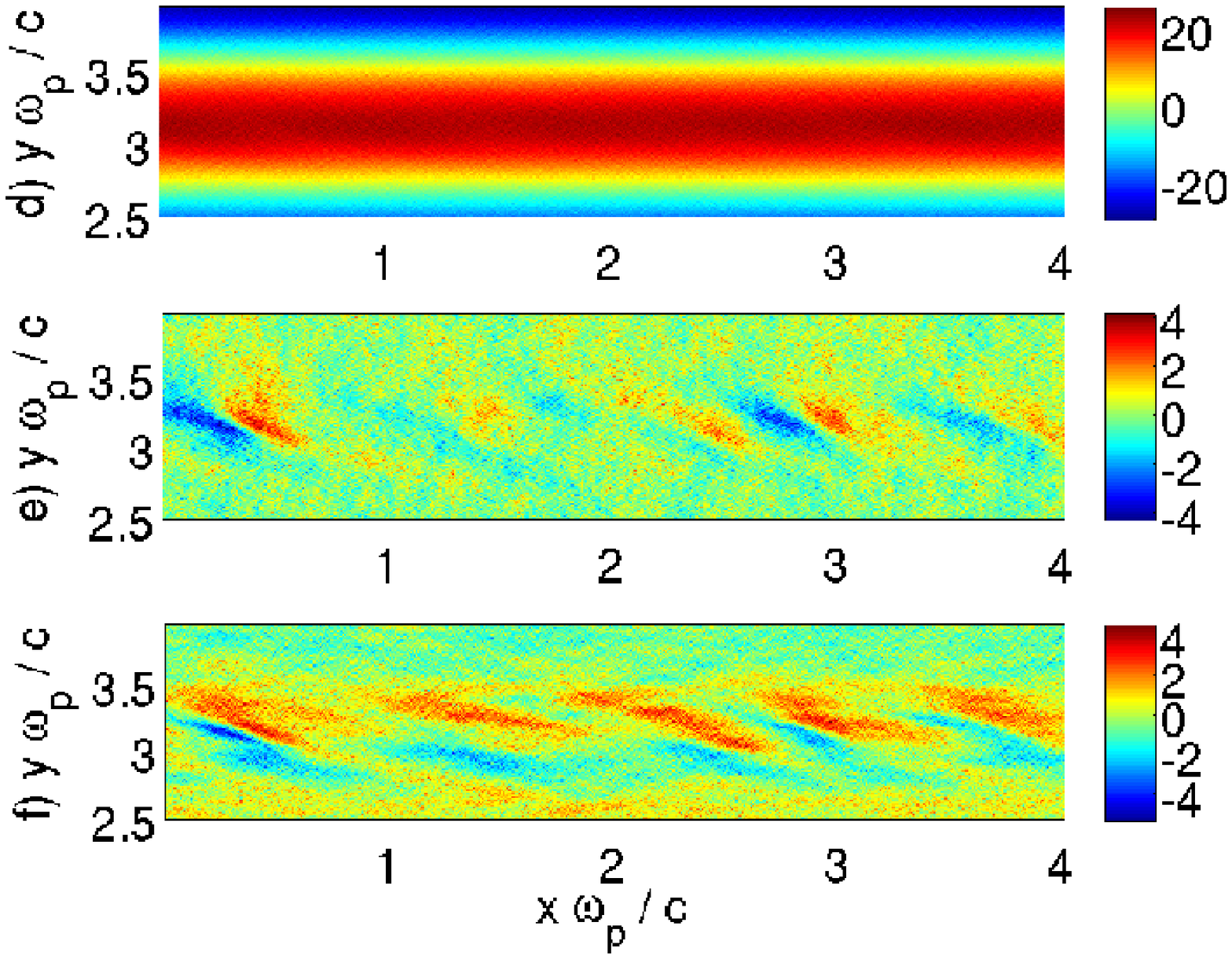} \caption{The relevant
	field components at two times in a box sub-interval in units
	of V/m (colours online): The \( cB_z \), \( E_x \) and \( E_y
	\) components at \( t \omega_p = 75 \) are shown in the panels
	a,b,c, respectively. The panels d,e,f display the \( cB_z \),
	\( E_x \) and \( E_y \) components at \( t \omega_p = 87 \),
	respectively.}\label{fig:NewFig}
\end{figure}

The mechanism that transfers the electric field energy density from the
\( E_y \) to the \( E_x \) component is revealed by comparing the spatial
field profiles before and after the growth of \( E_x \) in the Fig.
\ref{fig:NewFig}. 
The \(E_y\)-component is practically uniform along \(x\) and the structures
are parallel to those of \(cB_z\) at \( t \omega_p = 75 \) and both are
likely to be correlated. The amplitude \(E_y\) performs a full oscillation
in the spatial interval \( 2.6 < y \omega_p / c < 3.6\) between two
zero-crossings of \(c B_z\). No significant structures can be seen in 
\( E_x \) at this time. The \(c B_z \) component is qualitatively unchanged 
in the time interval \( 75 < t\omega_p < 87 \) also on this scale, but the 
oscillation amplitude along \( y \) decreases somewhat. The electric field 
does, however, undergo a drastic change. The \(E_y\) that is initially 
uniform along \( x \) becomes oscillatory at the later time. One oscillation 
takes place on a distance \( 0.8 c/\omega_p \) along \( x \), which
corresponds to the wavenumber $k_x c / \omega_p \approx 8$, as can be 
seen best from the \( E_y\) component. The oscillations along \( x \) rotate 
the electric field polarization vector away from \( y \) at \( t \omega_p = 
87 \). 

The observed spatial periodicity along \( x \) is evidence for a
secondary instability, possibly a sausage mode instability
\cite{Sausage}. It is this instability that couples the energy from
the \( E_y \) component into the \( E_x \) component (see
Fig.\ \ref{fig:S1enden}). This is confirmed by the supplementary movie
1, which animates in time the 10-logarithmic spatial power spectrum \(
{|\tilde{E} (k_x,k_y)|}^2 \) of \( \tilde{E}(x,y) = E_x(x,y) +
iE_y(x,y) \). The movie 1 demonstrates that the electric field power
is concentrated at $k_x \approx 0$ until $t \omega_p \approx 80$,
which is equivalent to a spatially uniform electric field along \( x
\). A signal grows after $t\omega_p \approx 80$ at $k_x c / \omega_p
\approx 8$, which corresponds to the oscillation of \( E_y \) in the
Fig.\ \ref{fig:NewFig}(f). This signal is spread over a wide band of
\( k_y \). Initially this signal grows only in the quadrant \( k_x,
k_y > 0 \), but it rapidly spreads out in the wavenumber space after
\( t \omega_p \approx 100 \). In \cite{mor} this secondary instability
was not observed as only the magnetic field has been analyzed.

\subsection{The electron phase space distribution}

The variation of \( B_z \) along \( y \) must be supported by a
current component \( J_x \) that varies along \( y \). In order to
obtain information about the distribution of the electrons in terms of
the normalised momentum \(p_x/m_e c\), we slice out 6 cells (\(0.084
\leq x \omega_p / c \leq 0.134\)) in the \(x\)-direction and integrate
the phase space density over this \( x \) interval to improve the
particle statistics.  We compare the electron phase space distribution
\( f(y,p_x)\) to \( J_x (y) \).

\begin{figure}[!ht]
	\centering \includegraphics[width=7cm]{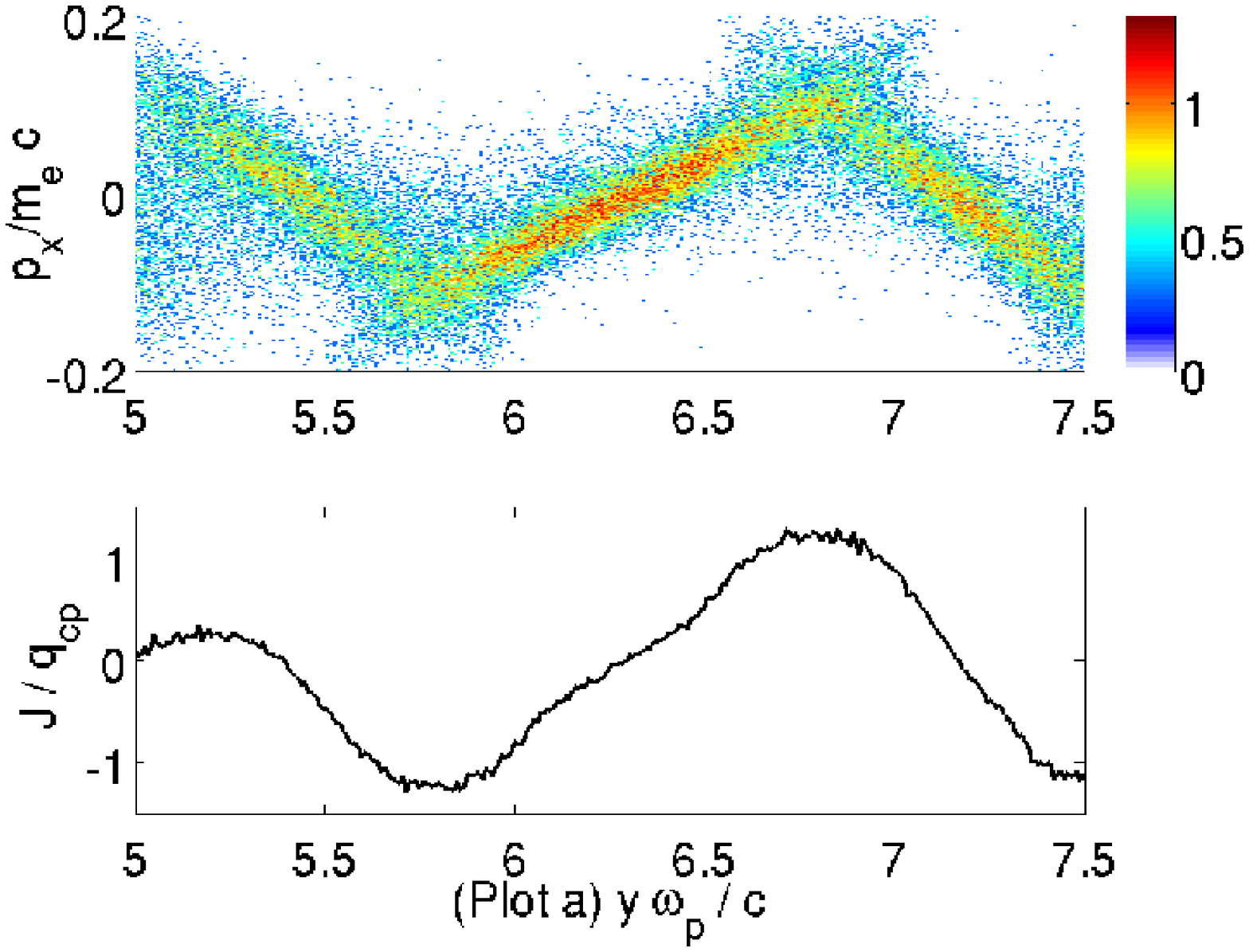}
	\includegraphics[width=7cm]{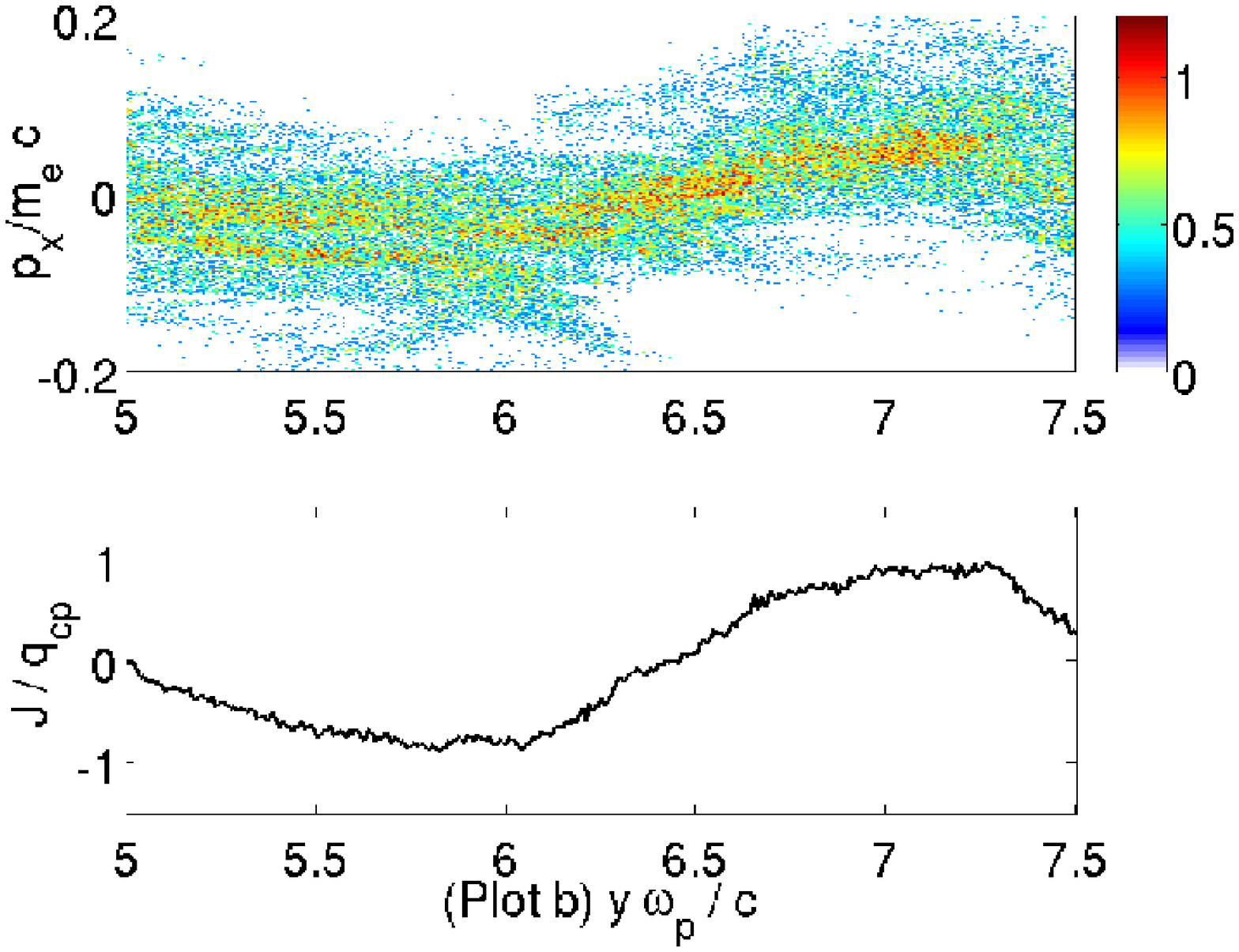} \caption{{\it Top:}
	The 10-logarithmic momentum distribution \(f(y,p_x)\) at the
	simulation times \(t \omega_p = 59\) (left) and \(t \omega_p =
	164\) (right) (colours online): The structure of the electron
	phase space distribution is initially an approximately
	piecewise linear function of \( y \). The structure is diffuse
	at late times. {\it Bottom:} \(J_x(y)/ q_{cp}\) at \( t
	\omega_p = 59 \) (left) and \( t \omega_p = 164 \)
	(right). The current initially varies piecewise linearly with
	\( y \). It is smoothened at late times and its
	amplitude has decreased.}  \label{fig:vel}
\end{figure}

Fig.\ \ref{fig:vel} displays \( f (y, p_x)\) and \( J_x (y) \) for the
two simulation times corresponding to Fig. \ref{fig:S1BZ120}. At \(t
\omega_p = 59\) the mean momentum of the bulk electrons changes
piecewise linearly with \( y\) and it spans a range \(|p_x / m_e|
\apprle v_{th\parallel}\). This results in a piecewise change of the
current \( J_x \). The \( B_z (y) \) is consequently varying quadratic
over extended \( y \) intervals (not shown). At \(t \omega_p = 164\)
these phase space structures become more diffuse. The magnetic field
amplitude in Fig.\ \ref{fig:S1BZ120} decreases by the reduction of
\(J_x\) at late times and, consequently, the energy density of \(B_z\)
in the Fig.\ \ref{fig:S1enden}(a).

\section{One-dimensional simulation}\label{einD}

We examine now with such a 1D simulation the time beyond \( t\omega_p
\approx 50\), when the TAWI grows the \( E_y \) field. The
one-dimensional geometry suppresses \(E_x\).

\begin{figure}[!ht]
	\centering \includegraphics[width=5cm]{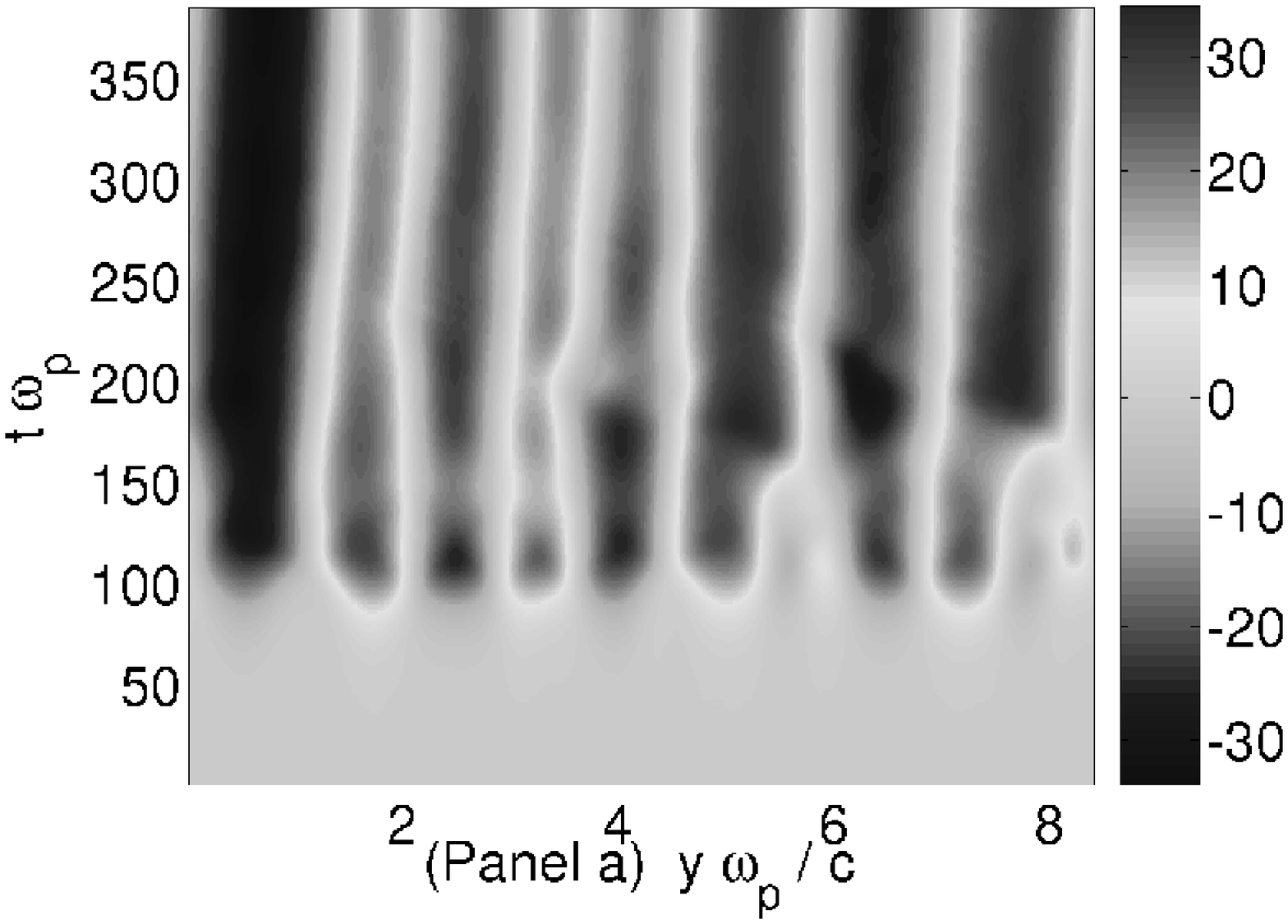}
	\includegraphics[width=5cm]{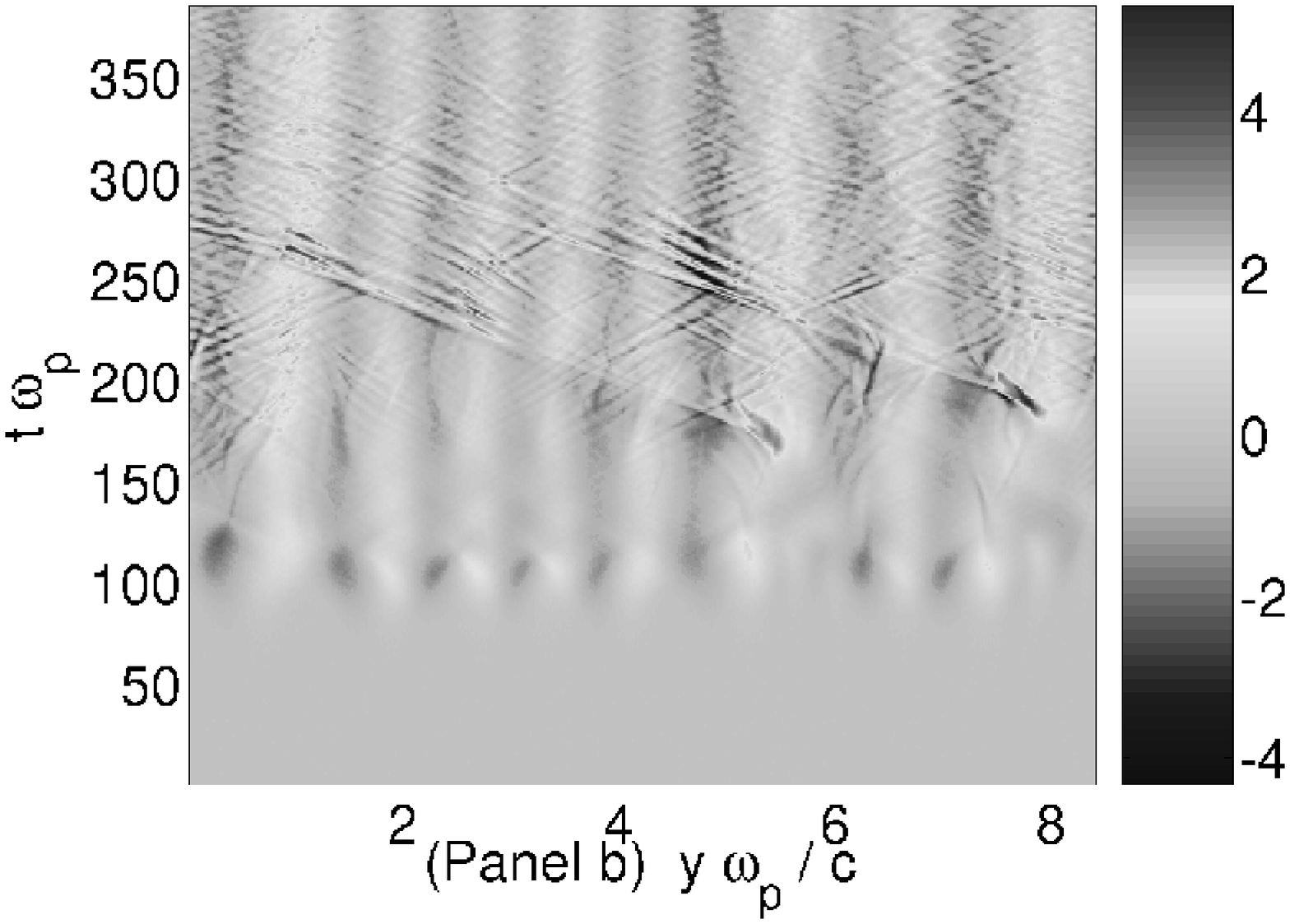}
	\includegraphics[width=5cm]{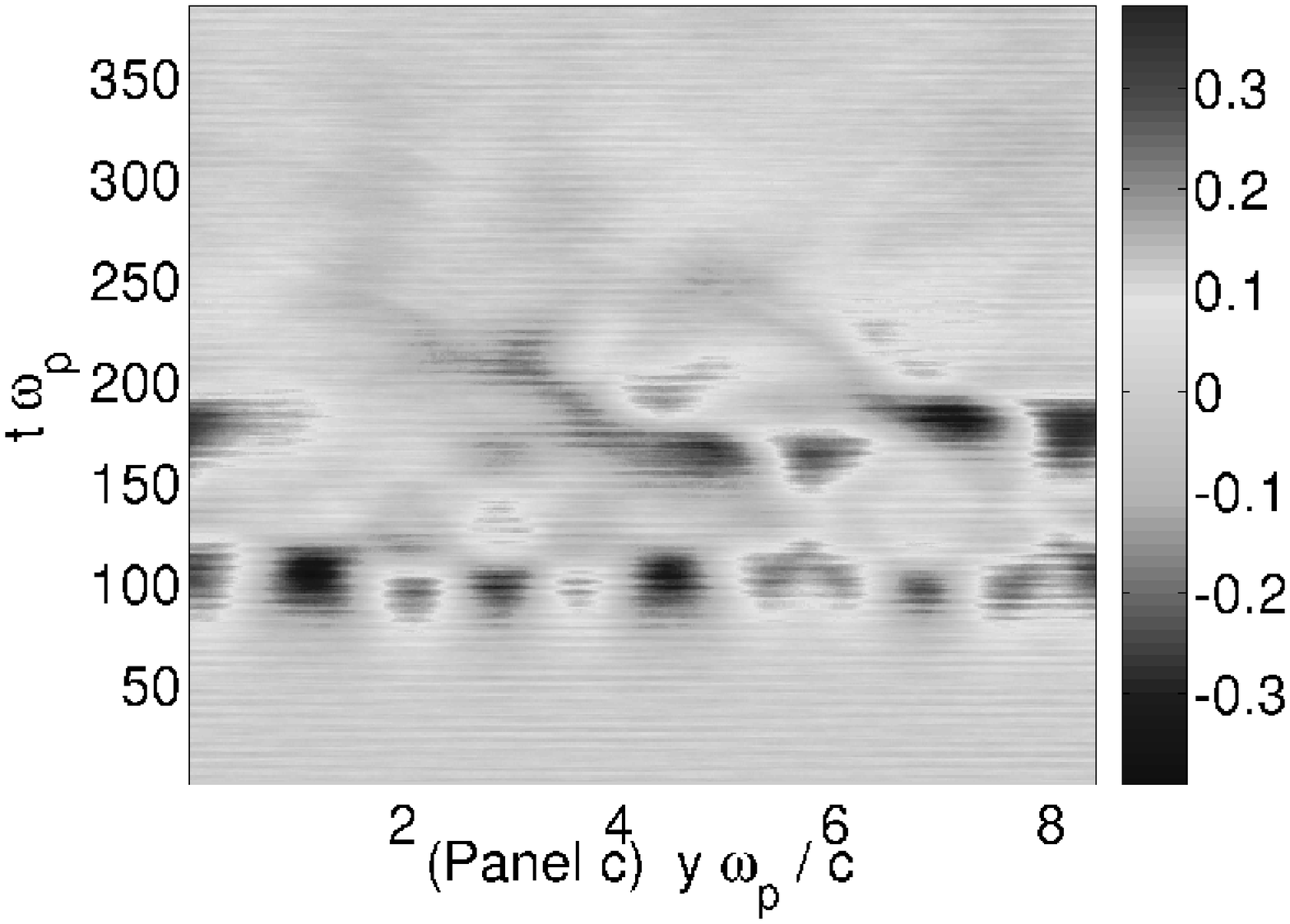}\caption{The development of
	the field components \(cB_z\), \(E_y\) and \(E_x\) during the
	simulation. The colourbar denotes the value in V\(/\)m. (a):
	The magnetic field structure is quasi-static and oscillatory
	in space. (b): \(E_y\) has a double frequency compared with
	\(cB_z\). Fast waves appear for \(t \omega_p > 180\). (c): The
	low frequency waves of \(E_x\) are generated by Ampere's law
	with a resulting phase shift of \(90^\circ\) towards
	\(B_z\).}\label{fig:1Dfields}
\end{figure}

As in the 2D simulation only the \(z\)-component of the magnetic field
has developed. The magnetic field structures are quasi-static and
oscillatory in space (Fig.\ \ref{fig:1Dfields}(a)). The structure of
\(E_y\) in Fig.\ \ref{fig:1Dfields}(b) has the half wavelength of
\(cB_z\) like in Fig.\ \ref{fig:S1BZ120}(a). For \(90 \leq t\omega_p
\leq 180\) the structure is time-independent. This stationarity is an
artifact of the 1D geometry. It permits us to separate the effects of
\( E_x \) and \( E_y \) on the electrons, which simplifies their
interpretation. For \(t\omega_p > 180\) fast waves appear in \( E_y
\), which are superposed on the periodic structure. Their velocity \(v
\approx c/20\) is of the order of \(v_{th\parallel}\).

The \(E_x\) in Fig.\ \ref{fig:1Dfields}(c) resembles \(cB_z\) for \(t
\omega_p < 120\) and has an amplitude that is one order of magnitude
below that of \(E_y\). Eq.\ (\ref{thirteen}), which is in one
dimension \(\partial_y B_z = \partial_t E_x + J_x\), implies that a
growing \( J_x \) generates the oscillations of \(E_x\). The phase
shift of \(90^\circ\) between \(cB_z\) and \(E_x\) shown in
Fig.\ \ref{fig:1Dfields} is an evidence for this connection. The
oscillations of \( E_x \) correspond to transient waves and they damp
out as evidenced also by Fig.\ \ref{fig:S1enden}(b).

Movie 2 shows the evolution of the electron charge density and the
field components \(c B_z\) and \(E_y\). Initially, the growth of the
\(cB_z\) and \(E_y\) components reveals a spatial correlation and the
electron charge density is modulated accordingly.  Oscillations of the
electron charge density and of the \(E_y\) component with a short
wavelength develop at late times. These rapid spatial oscillations
obscure the long spatial oscillations. The latter are, however, still
present, as the Fig.\ \ref{fig:1Dfields} demonstrates for \(E_y\). The
oscillation of \(B_z\) with the initial wavenumber remains dominant.

\begin{figure}[!ht]
	\centering \includegraphics[width=5cm]{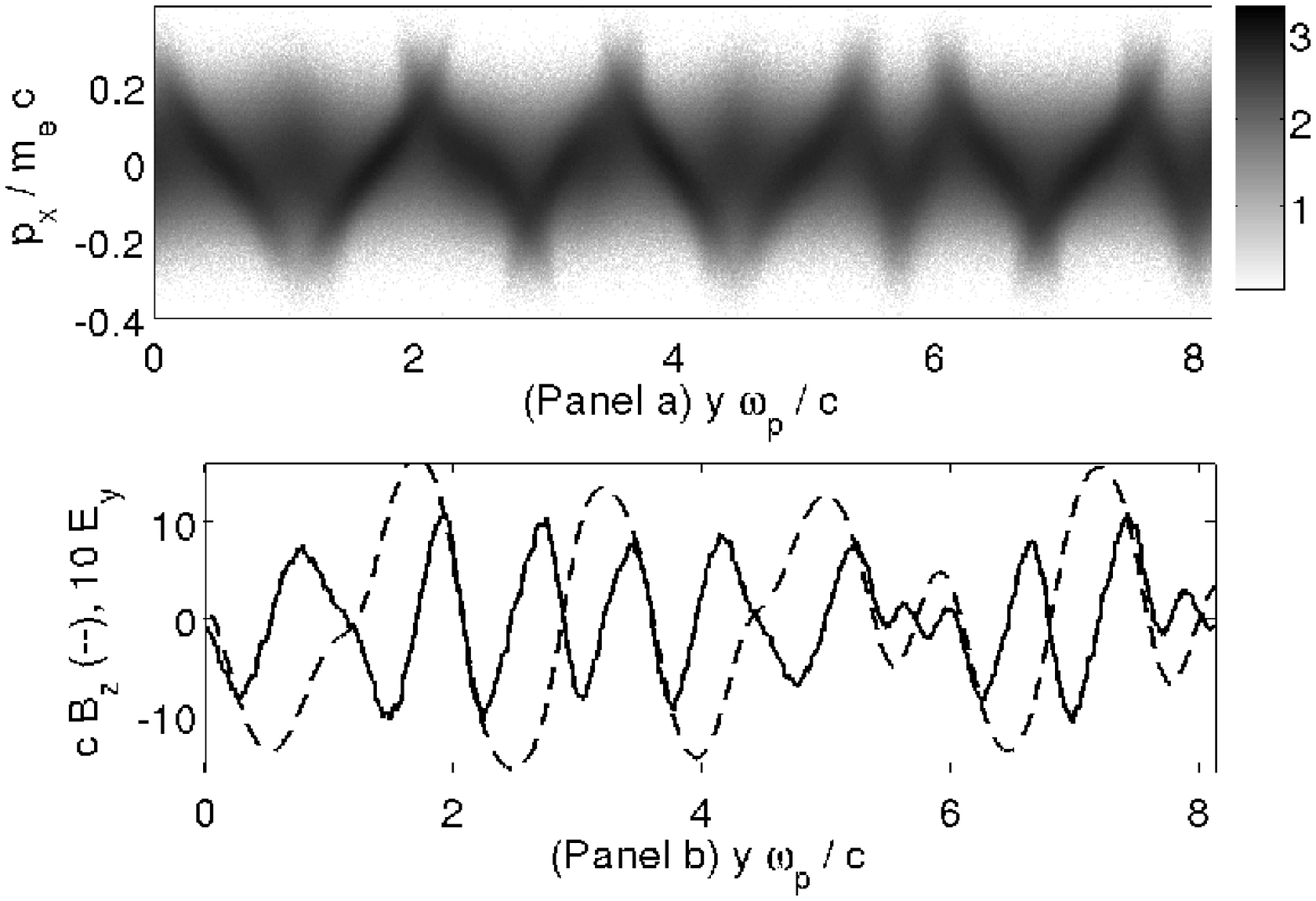}
	\includegraphics[width=5cm]{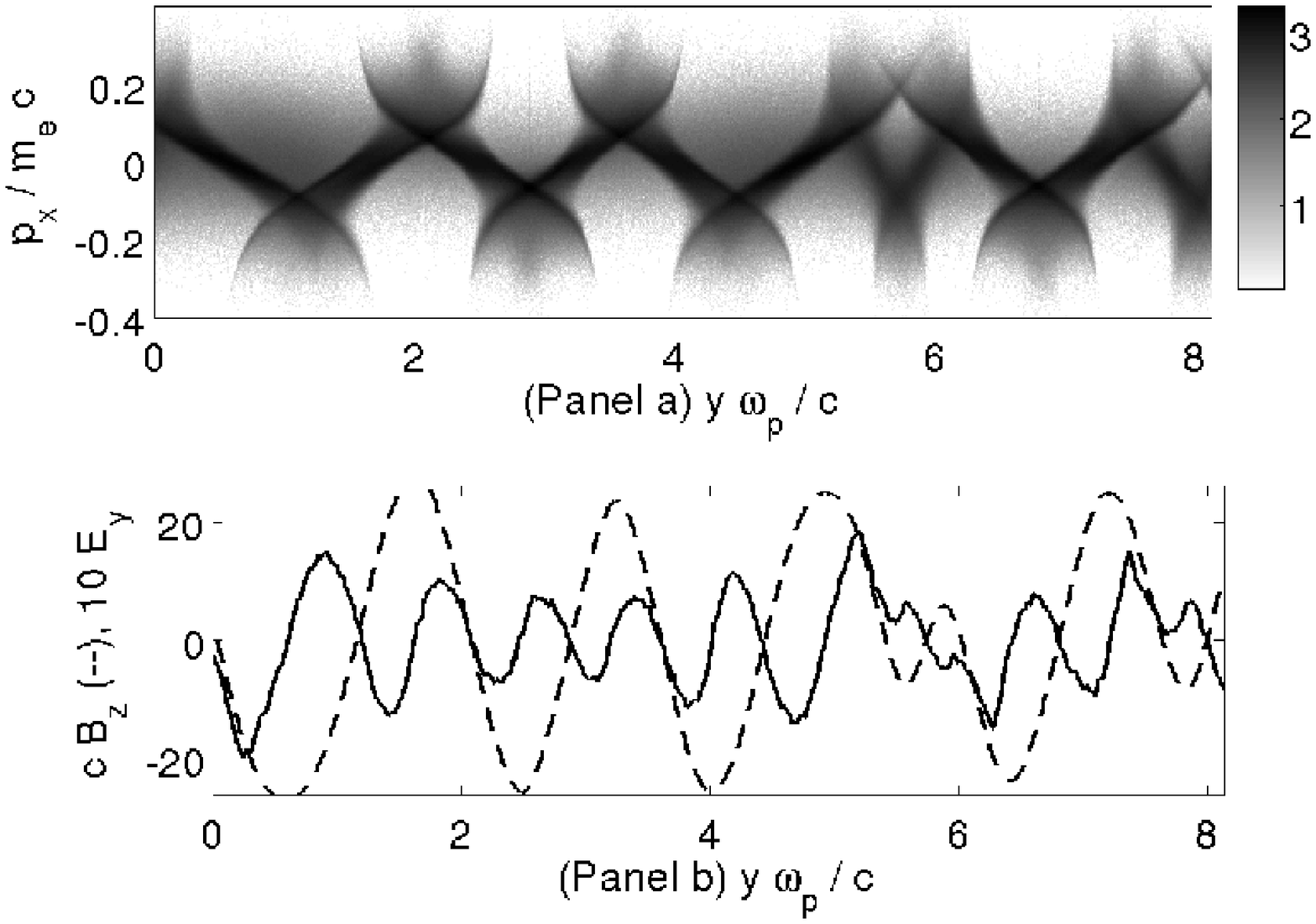}
	\includegraphics[width=5cm]{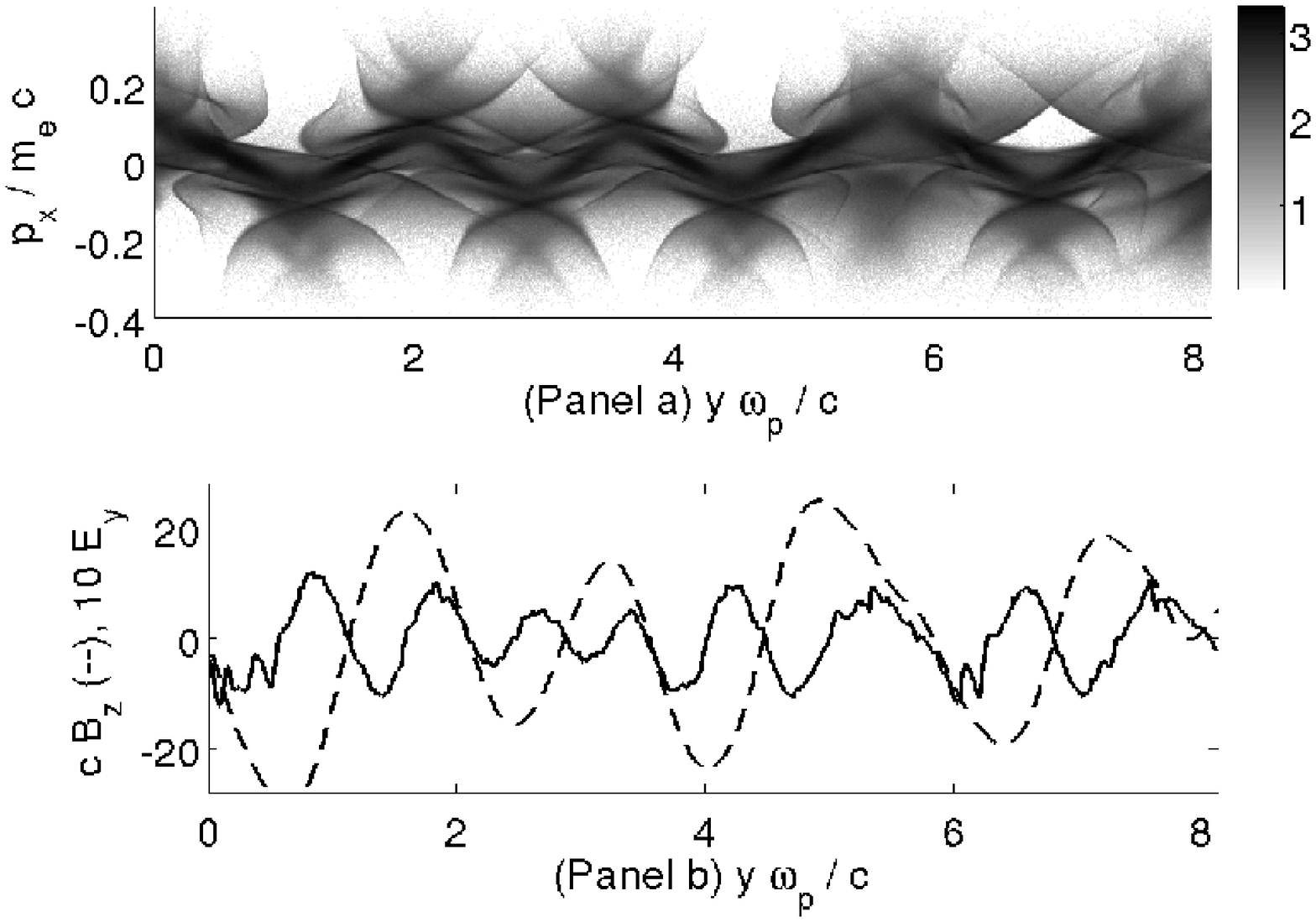}
\caption{Row a: \(f(y,p_x)\) of the particles at \(t \omega_p = 96\)
  (left), \(t\omega_p = 116\) (middle) and \( t \omega_p = 154\)
  (right). The colourbar denotes the 10-logarithm of the number of
  CPs. Row b: \(cB_z\) (dashed) and \(10 E_y\) (solid) at the times
  corresponding to the electron distributions above. The \(y\)-axis
  denotes the value in V/m.}\label{fig:PhaseFeld506080}
\end{figure}

In Fig.\ \ref{fig:PhaseFeld506080} snapshots are taken and the
connection with the phase space distribution is investigated. In the
1D simulation the electrons can move freely only along the
\(y\)-direction. The TAWI generates initially a zigzag distribution,
which yields through its currents the growth of the magnetic
field. The amplitude of the velocity oscillations upon saturation are
comparable to \( v_{th\parallel} \), which is clearly recognizable at
\(t \omega_p =96 \) (See left column of
Fig.\ \ref{fig:PhaseFeld506080} and movie 3). The wave number of
\(E_y\) is twice that of \(c B_z\) at this time and the zero crossings
of \(E_y\) coincide with the extrema and zero crossings of
\(cB_z\). At this stage the magnetic field still grows
exponentially. Forces due to \( E_y \) and \( v_{cp}B_z\) are not yet
strong enough to affect the electron trajectories. Until strong
electromagnetic fields have developed, the electrons perform just a
thermal motion in the \(y\)-direction. Once the electromagnetic fields
become strong at \( t \omega_p = 116 \), they push together the
particles along the \(y\)-direction, yielding a layered structure in
the phase space distribution (middle column
Fig.\ \ref{fig:PhaseFeld506080}). The complexity of the phase space
distribution increases further with the time, which is evident from
the movie 3 and the right column in
Fig.\ \ref{fig:PhaseFeld506080}. The fine structure at the beginning
of the simulation thermalizes but the zigzag distribution of the bulk
remains. This holds up the currents which are neccessary to support
the magnetic field. A coalescence of the filaments is not apparent.

\begin{figure}[!ht]
	\centering \includegraphics[width=7cm]{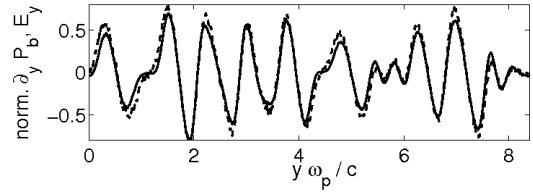} \caption{The
          normalized magnetic pressure gradient \(2\partial_y P_b/ n_e
          e\) (solid) and the negative electric field component
          \(-E_y\) (dashed) in values of V/m at \(t \omega_p =
          96\).}\label{fig:1Dprgrad}
\end{figure}

In order to demonstrate quantitatively the link between the magnetic
pressure gradient and \( E_y \), these are plotted in
Fig.\ \ref{fig:1Dprgrad}(a). The expected proportionality of the
magnetic pressure gradient and the electric field
\begin{equation}
	-e E_y(y) = n_e^{-1} \, \partial_y P_b(y) = n_e^{-1} \mu_0^{-1} B_z(y) \, \partial_y B_z(y)
\end{equation}
had to be modified by a factor 2. This factor can be understood as
follows. The magnetic pressure gradient force accelerates the
electrons, giving rise to a $J_y \neq 0$. The $J_y$ is tied through
Amp\`ere's law to an $E_y$ and the phases of both are shifted in time
by \(90^\circ\). Initially, $E_y = 0$ and $J_y = 0$ and $E_y$ must
thus oscillate around an electric field set. The peak value of $E_y$
is given by the added fields of the oscillatory $E_y$ and of this
set. This amounts to twice the amplitude of the set electric field,
which is determined by the magnetic pressure gradient force.

\begin{figure}[!ht]
	\centering \includegraphics[width=7cm]{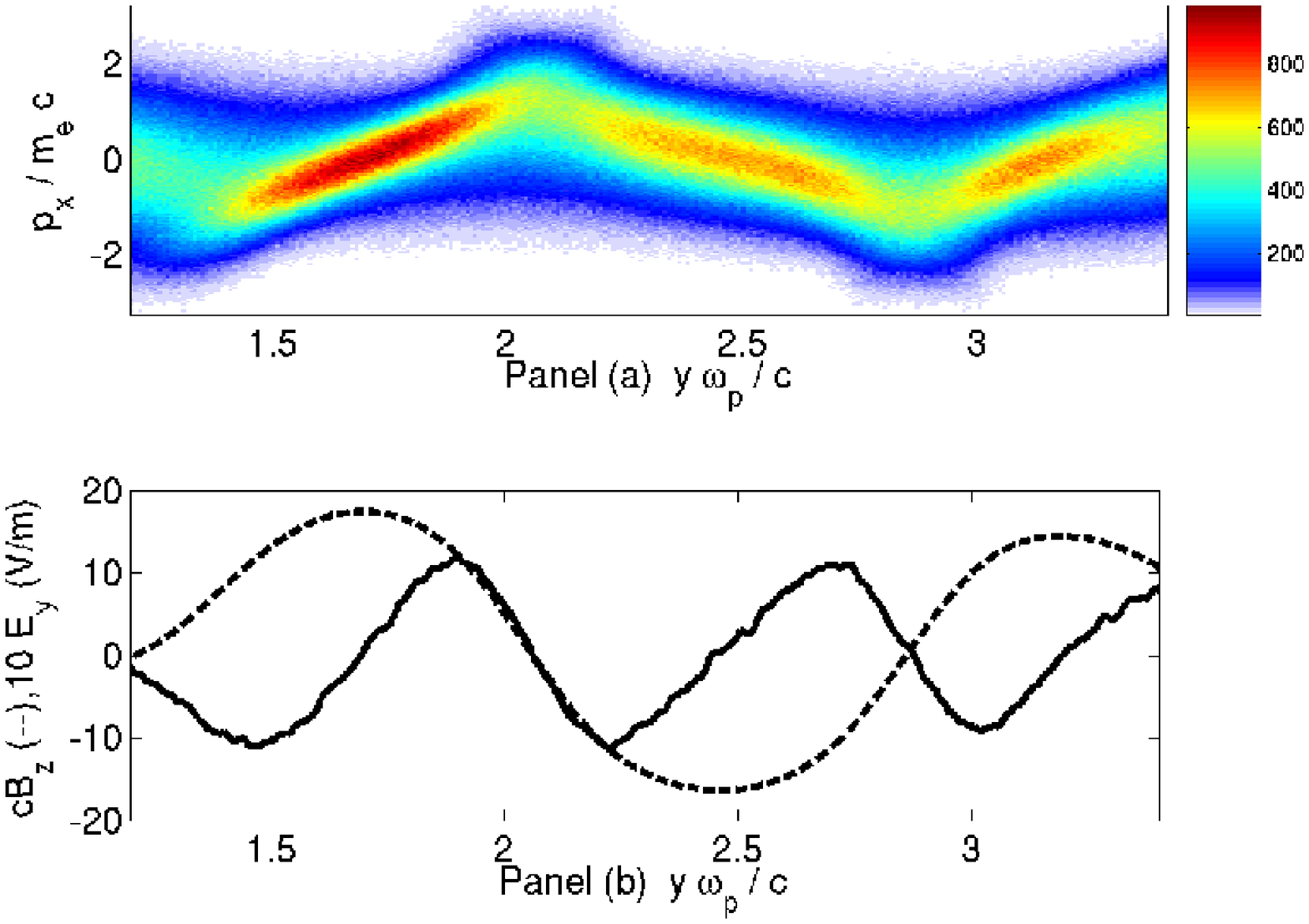}
        \includegraphics[width=7cm]{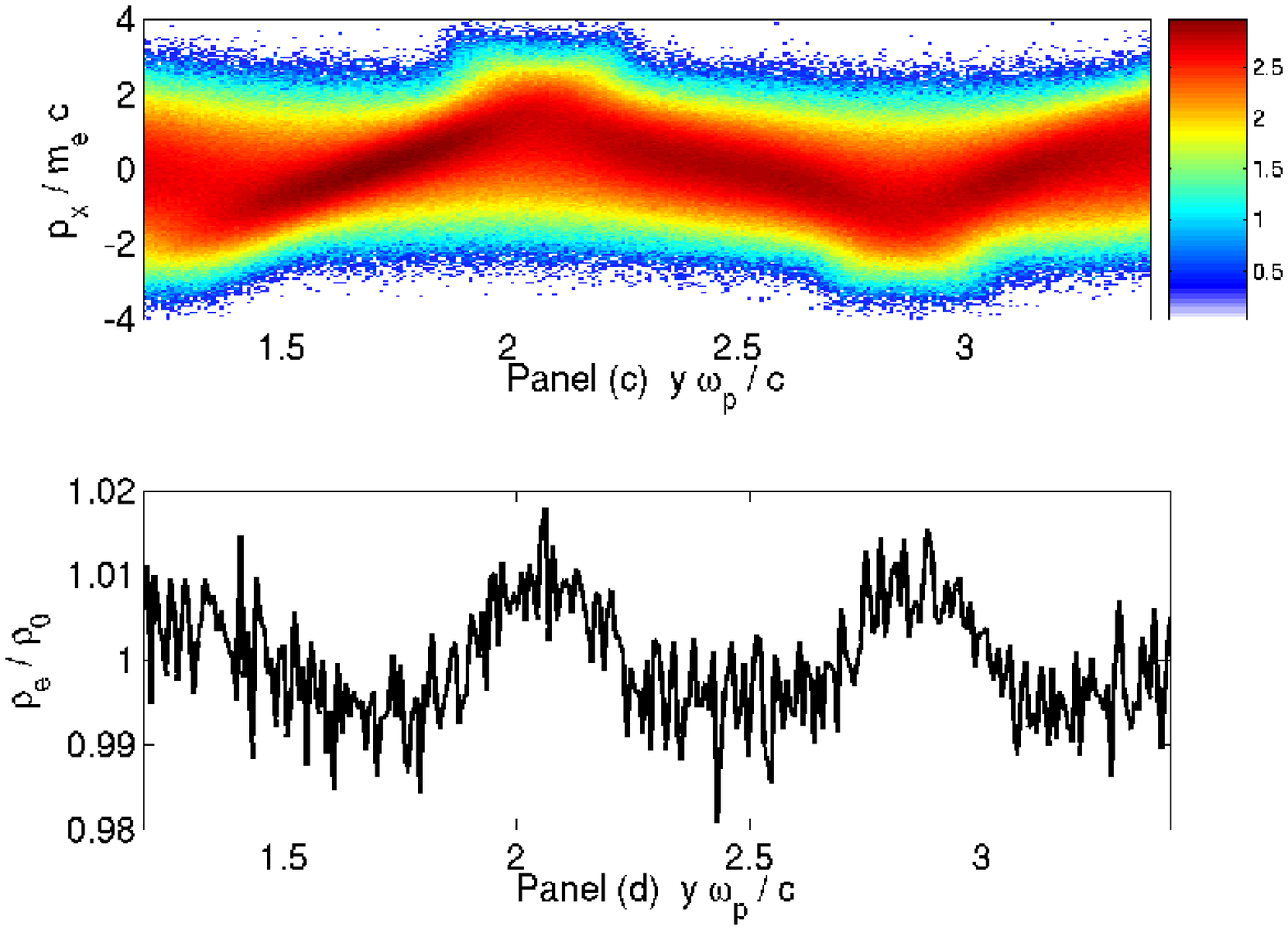} \caption{The electron
          distribution and the electromagnetic fields in a
          sub-interval of the box at \(t\omega_p = 96\) (colours
          online). (a) The electron phase space density in units of a
          CP on a linear scale. The density decreases from the center
          outwards. (b) The electrostatic field \(10 E_y \) (solid) and
          the \(cB_z \) (dashed) due to the TAWI. (c) The
          10-logarithmic electron phase space density in units of a
          CP. (d) The electron charge density normalized to its mean
          value.}\label{phaseshift}
\end{figure}

Fig.\ \ref{phaseshift} analyses the electron distribution and the
fields in a subsection of the simulation box at the time \( t \omega_
p = 96 \). The electron phase space distribution follows the zigzag
distribution. The width of the distribution in \( p_x \) with more
than 300 particle counts is practically constant along \( y \) on the
linear and 10-logarithmic density scales. The thermal spread and the
electron density increase somewhat at the breaks, which coincide with
the zero-crossings of \( E_y \) and \( c B_z \).

A comparison of the Figs.\ \ref{phaseshift}(a,b,d) reveals the cause
of the density accumulation and the zigzag distribution. Let us
consider the interval to the right of the break at \( y \omega_p / c
\approx 2.1 \).  There, the electron bulk has a \( p_x > 0 \) and the
\( E_y, B_z < 0 \).  The \( E_y < 0 \) implies that the electrons
accelerate away from the break in the direction of increasing \( y
\). The electrons accelerate away from the break also for the interval
just to the left of \( y \omega_p / c = 2.1 \), where \( E_y > 0
\). The break is thus an unstable equilibrium in what the \( E_y \)
driven by \( \nabla P_b \) is concerned and it cannot explain the
electron accumulation close to \( y \omega_p / c = 2.1\) in Fig.
\ref{phaseshift}(d). This is accomplished by the \( B_z \). The
Lorentz force component along \( y \) due to \( B_z \) for a
computational electron with \( q_{cp} < 0, v_{cp,x} > 0, B_z < 0 \)
just to the right of \( y \omega_p / c = 2.1 \) is \( F_{cp} = -q_{cp}
v_{cp,x} B_z < 0 \) and the electron is accelerated towards \(
y\omega_p / c = 2.1 \). The particles are also accelerated towards the
break just to the left of it, because \( B_z \) changes its sign. It
is thus the drift electric field \( \propto v_x B_z \) that compresses
the electrons. The electrostatic force that is due to the magnetic
pressure gradient, which oscillates twice as fast as \( B_z \) along
\( y \), opposes this compression close to the break and enforces it
further away. The force contribution by the electrostatic field is not
negligible.  The peak speed along $x$ of the electrons is $0.1c$ and
the Lorentz force $v \times B$ is thus comparable to that due to
$E_y$.

\section{Discussion}\label{discussion}

In this work we have examined the thermal anisotropy-driven Weibel 
instability (TAWI). We have limited our investigation to an unmagnetized, 
non-relativistic plasma, considering only the electron dynamics and 
neglecting the ions. The TAWI is driven by a temperature anisotropy in 
the electron distribution. Here, the direction of higher temperature has 
been aligned with the \(x\)-direction. Thus, the TAWI modes would have 
wave vectors in the plane orthogonal to \(x\).

We have performed two simulations. The first one is a 2D simulation in
the \(x\)-\(y\)-plane. As the electrons can move freely only in this
plane, the waves that grow due to the TAWI are planar with a wave
vector along \(y\). Second we have performed a 1D simulation along
\(y\) with a large number of computational particles per cell. This
has enabled through an increased dynamical range the identification of
fine structure in the electron phase space distribution and it has
reduced the noise. Both simulations have not resolved the \( y\)-\(z
\)-plane, which would be necessary to enable the interplay of the
current filaments \cite{mor}. We can thus separate the dynamics of
individual current filaments driven by the TAWI from their mutual
nonlinear interactions, as in \cite{mor}, and leave the latter to
future work.

As expected, only the \(z\)-component of the magnetic energy density has 
been amplified \cite{mor}. After an initial exponential growth phase of
\( B_z \), the wave has saturated. The growth of \( B_z \) has been
accomplished by the turning of the initially spatially uniform electron
velocity distribution into a zigzag distribution in the phase space. The
mean speed along \( x \) varies for this distribution piecewise linearly 
as a function of \( y \). The initial \( J_x (y) = 0 \) is thus transformed
into a \( J_x (y) \neq 0 \) in most parts of the simulation box. In both 
simulations the saturation resulted in the growth of an electrostatic 
\( E_y \) component. We have demonstrated with the 1D simulation the link 
between this \( E_y \) and the force due to the magnetic pressure gradient 
\( \nabla P_b \). This connection was qualitatively unchanged in the 2D 
simulation, but the box-averaged electrostatic energy density did not grow 
at twice the exponential rate of the \( B_z \), as in the 1D simulation of 
the TAWI and in simulations of the filamentation instability 
\cite{cal,die7,sto3}. The \( E_x \) component could grow to a high amplitude 
only in the 2D simulation. The \( E_x \) in the 1D simulation has been
driven by the growing \( J_x \) according to Ampere's law but only to a low
amplitude. These modes damped out rapidly. The stronger \( E_x \) in the 
2D simulation developed in response to a secondary instability, probably a 
sausage instability \cite{Sausage}. The 1D simulation cannot resolve this 
instability, because it involves wavevectors along \( x \).
 
The 1D PIC simulation provided an insight into the link between the
fields and the electron phase space structures and it could resolve
the charge density oscillations that developed during the nonlinear
stage of the TAWI. The breaks in the zigzag distribution have been
examined in more detail. The breaks correspond to the positions, at
which the change with \( y \) of the mean momentum along \( x \)
reverses its sign. The charge density and the electron temperature are
increased at the breaks and the amplitudes of \( E_y , B_y \) have
zero-crossings. The \( E_y \) field accelerates electrons away from
the break, while the drift electric field imposed by the non-zero mean
speed along \( x \) and the \( B_z \) is compressing the electrons
along \( y \). This compression is stronger close to the breaks than
the electric repulsion, which results in an electron accumulation.

We have demonstrated that the force due to the electric field is
comparable to the Lorentz force imposed by the magnetic field, because
the electron speeds are only a few per cent of the light speed. The
focus on the magnetic forces and the neglection of the electric ones
in Ref.\ \cite{mor} may thus not always be justified for the Weibel
instability. The high phase space resolution of PIC simulations, which
is now possible, has revealed that the size of the filaments in the 1D
simulation increases not through the merger of filaments, which is
inhibited by the 1D geometry, but through a broadening of the
filaments due to the overlap of phase space layers. The filament
mergers suggested by Fig.\ 3 in Ref.\ \cite{mor} could be an artifact
of the lower plasma resolution or from the display of phase space
slices instead of the full distribution. This has to be investigated
with a future simulation that employs the anisotropy \(A=24\) of
Ref.\ \cite{mor}.

The connection of the magnetic and electric fields driven by the TAWI
thus resembles those reported for the filamentation instability, which
is driven by counterpropagating electron beams. Stable plasmons
developed in \cite{die7} after the filamentation instability has
saturated, which were confined by the potential due to the drift
electric field and the magnetic pressure gradient. Here, the electron
phase space distribution in the 1D simulation developed fine
structures first evolving into a thermalized distribution with a bulk
zigzag distribution.

We will investigate in future work the statistical properties of the
TAWI in 1D and in 2D, as it has been done for the filamentation
instability \cite{row,die6}. Ultimately, a 3D simulation has to be
performed as in Ref.\ \cite{rom}, which considered the original Weibel
instability in which the electrons are hot along two cartesian
directions and cool along the third. These simulations may provide an
insight into the scale size and field strength of the magnetic field
structures that develop out of nonthermal electron distributions in
astrophysical flows and to what extent these structures could be
responsible for the frequently observed synchrotron radiation.

{\bf Acknowledgements:} This work was partially supported by the
Forschergruppe 1048 of the Deutsche Forschungsgemeinschaft through
grant Schl201/21-1. ME Dieckmann has been financed by Vetenskapsr\aa
det and through the grant FOR1048 of the Deutsche
Forschungsgemeinschaft. We thank the HPC2N supercomputer centre for
the computer time and support. The Workshop ``PIC Simulations of
Relativistic Collisionless Shocks'' in Dublin, Ireland (May 19 -- 23,
2008) provided us with helpful suggestions.

\end{document}